\documentclass{ws-procs975x65}
\usepackage{float}
\usepackage{graphicx}

\def\beq{\begin{equation}}
\def\eeq{\end{equation}}
\def\rmd{\mathrm{d}}
\begin{document}

\title{The Struble--Einstein Correspondence}
\author{Marcus C. Werner}

\address{Center for Gravitational Physics, Yukawa Institute for Theoretical Physics, \\ Hakubi Center for Advanced Research, \\ Kyoto University,\\
Kitashirakawa Oiwakecho Sakyoku, Kyoto 606-8502, Japan \\
E-mail: werner@yukawa.kyoto-u.ac.jp}

\begin{abstract}
This article presents hitherto unpublished correspondence of Struble in 1947 with Menger, Chandrasekhar, and eventually Einstein, about a possible observational test supporting Einstein's special relativity against Ritz's emission theory using binary stars. This `Struble effect,' an acceleration Doppler effect in emission theory, appears to have been overlooked, and the historical context, including de Sitter's binary star test of special relativity, is also discussed.  
\end{abstract}


\bodymatter

%%%%%%%%%%%%%%%%% now a standard article style for the most part

\section{Introduction}
The nature of light propagation is central to the history of relativity theory. An early rival of Einstein's special relativity was emission theory, which was also consistent with the negative result of the Michelson-Morley experiment. In 1947, Raimond Struble (1924--2013), then a student at the University of Notre Dame and later a professor of mathematics at North Carolina State University in Raleigh, corresponded with Chandrasekhar at Yerkes Observatory and Einstein at the Institute for Advanced Study about an observational test of emission theory using binary stars. While de Sitter, among others, had shown earlier that binary star observations do, indeed, support special relativity against emission theory, Struble pointed out a new Doppler effect due to the binary's orbital acceleration, which appeared to have been overlooked.

These letters and related documents, hereinafter denoted \textit{Struble-Einstein correspondence} (SEC), are held privately and are listed in the Appendix. The purpose of this article is twofold: on the one hand, to present this unpublished correspondence with Einstein and, on the other hand, to draw attention to Struble's effect and place it in its historical context of binary star tests of emission theory.

\section{Historical background}
\subsection{Emission theory}
After Einstein's paper on special relativity\cite{einstein1} was published in 1905, emission theory was considered as a viable alternative to understand the negative result\cite{michelsonmorley} of the Michelson-Morley interferometer experiment without the need to adopt special relativistic kinematics. This theory was developed in particular by Walter Ritz (1878--1909), Einstein's junior by one year at the Zurich Polytechnic Institute (presently the ETH), who proposed such a framework\cite{ritz} in 1908 (see, e.g., Ref.~\refcite{martinez} for historical aspects): light is emitted at a constant speed $c$ in any direction $\mathbf{u},\ |\mathbf{u}|=1,$ relative to a source moving at velocity $\mathbf{v}$ relative to an observer. Thus, the velocity of light relative to the observer becomes
\begin{equation}
\mathbf{c}_r=c\mathbf{u}+\mathbf{v}\,.
\label{ritz}
\end{equation}
Just as Einstein's theory, this is consistent with Michelson-Morley because light has a constant speed relative to the interferometer in this case (without being {\it universally} constant), so there is no diurnal variation of the light travel time.

To test the emission theory with a similar interferometer experiment, one would either require suitably moving mirrors, as was indeed considered by Michelson\cite{michelson} himself in 1913, or a light source moving fast with respect to the interferometer, such as a celestial light source. This point was made by Tolman in 1912:
\begin{quote}
``Hence if the Ritz theory should be true, using the sun as source of light we should find on rotating the apparatus a shift in the [interference] fringes of the same magnitude as originally predicted for the Michelson-Morley apparatus where a terrestrial source was used. If the Einstein theory should be true, we should find no shift in the fringes using any source of light.'' (Ref.~\refcite{tolman2})
\end{quote}
A negative result using star light was finally achieved by Tomaschek\cite{tomaschek} in 1924, providing strong evidence against emission theory and corroborating special relativity. Meanwhile, other experimental tests of emission theory were sought, and it is here that binary stars enter the picture.

\subsection{Binary star tests}
\label{sec:binary}
\begin{figure}
\centering
\includegraphics[width=12cm]{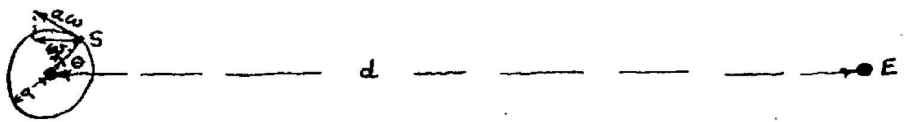}
\caption{Schematic binary star system as considered by Struble in SEC 3.}
\label{fig:binary}
\end{figure}

Consider an idealized Newtonian binary star system consisting of a small component $S$ orbiting a large component with speed $v$ in a circular orbit of radius $a$, whose centre is effectively the barycentre of the system. We assume also that the observer $E$ is coplanar with the orbit at large distance $d$, $d\gg a$, and that light is emitted isotropically by $S$ at the emission time $t_0$. Denoting the constant orbital angular velocity of $S$ by $\omega$, we have $v=a\omega$. This setup is illustrated in Fig.~\ref{fig:binary}.
\newline
Suppose, now, that emission theory holds. Then at emission time $t_0$, 
\begin{equation}
c_r(t_0)=c-v_r(t_0)\simeq c-v\cos \theta(t_0)=c-a\omega \cos \omega t_0\,,
\label{speed}
\end{equation}
from (\ref{ritz}). This speed is constant from $S$ to $E$, which are separated by the distance $d(t_0)\simeq d+a\sin \omega t_0$, so a light signal emitted by $S$ at $t_0$ is received by $E$ at the observation time
\begin{eqnarray}
t & = & t_0+\frac{d(t_0)}{c_r(t_0)}\simeq t_0 +\frac{d+a \sin \omega t_0}{c-a\omega \cos \omega t_0} \nonumber \\
\mbox{} & = & \frac{d}{c} + t_0 +\frac{a}{c}\sin \omega t_0+ \frac{ad\omega}{c^2}\cos \omega t_0 + \mathcal{O}\left(\frac{a^2\omega^2}{c^2}\right)
\label{time}
\end{eqnarray}
at leading order, with $v \ll c$. Hence, the observation time $t$ is a linear function of the emission time $t_0$ plus a small oscillation, as would be the case in Einstein's theory with $c_r(t_0)=c=\mathrm{const.}$ However, note that $d\omega /c$ can be large, so the fourth term in (\ref{time}) need not be negligible, and this gives rise to 
characterstic optical effects of emission theory: if $d\omega/c$ is sufficiently large, the map $t \mapsto t_0$ is no longer one-to-one, and a given observation time corresponds to multiple emission times. In other words, $E$ would observe $S$ at the same time in multiple points of its orbit. In his \textit{Introduction to the Theory of Relativity}, Bergmann referred to this phenomenon of multiple star images as ``ghost stars", whose absence provided strong evidence against the emission theory of Ritz and for the constancy of the speed of light in the sense of Einstein's special relativity theory:
\begin{quote}
``In some cases, we should observe the same component of the double star system simultaneously at different places, and these `ghost stars' would disappear and reappear in the course of their periodic motions. [...] However, no trace of any such effect has ever been observed. This is sufficiently conclusive to rule out further consideration of this hypothesis.'' (Ref. \refcite{bergmann}, pp. 19--20)
\end{quote}
Daniel Comstock (1883--1970), who visited J. J. Thomson at Cambridge during 1906--1907 before returning to MIT where he later became a co-founder of \textit{Technicolor} \cite{comstock3}, seems to have been the first to seriously consider astronomical tests of emission theory, in particular its implications for binary stars. At a meeting of the American Physical Society in October 1909 at Princeton, he noted that
\begin{quote}
``The assumption that the velocity of light depends on that of the source has, so far as the author is aware, never been properly examined. This is strange, but is probably explainable as a natural result of the complete trust which has been put for years in the conception of an ether.'' (Ref. \refcite{comstock1}) 
\end{quote}
In 1910, Comstock devised a criterion\cite{comstock2} to check whether the orbit of an observed binary obeys Keplerian motion, thus revealing anomalies such as a changing speed of light according to emission theory. Apparently unaware of Comstock's work, de Sitter also proposed\cite{desitter1, desitter2} to use binary star orbits to constrain emission theory in February 1913. This seems to have had more impact, and generated considerable interest within the physics community during that year, e.g. with papers by Guthnick\cite{guthnick} and Freundlich\cite{freundlich2}. In his reply\cite{desitter3}, de Sitter agreed to a parametric extension generalizing Eq. (\ref{ritz}),
\[
\mathbf{c}_r=c\mathbf{u}+\kappa\mathbf{v}\,,
\]
where $\kappa=1$ in Ritz's original theory and $\kappa=0$ in special relativity. The task, then, was to constrain $\kappa$, and by considering the system $\beta$ Aurigae, he found $\kappa<0.002$ and hence good agreement with special relativity. Einstein himself was, naturally, very interested in these developments. In May 1913, he wrote to Ehrenfest,
\begin{quote}
``The matter concerning binary stars is very nice, provided that the [spectral] line movements are really measured accurately enough to check the Keplerian motion to some extent.'' (Ref. \refcite{papers1}, doc. 441, p. 523)\footnote{Own translation. Original: ``Die Sache mit den Doppelsternen ist sehr h\"{u}bsch, vorausgesetzt, dass die Linienwanderungen wirklich exakt genug gemessen sind, um einigermassen die Keplersche Bewegung nachzupr\"{u}fen.''} 
\end{quote}
At this time, Einstein was working on the generalization of his theory of relativity to include gravity. Having predicted a gravitational deflection of light in 1911 using an early version of general relativity\cite{einstein2}, he was in touch with Freundlich who had proposed an observational test of gravitational lensing\cite{freundlich1} (cf. also Ref. \refcite{sauer}) in January 1913. In a letter to Freundlich in August 1913, mainly about this issue of gravitational lensing, Einstein also expressed the paramount importance of the binary star tests for relativity theory:    
\begin{quote}
``I am also very curious about the results of your investigations concerning the binary stars. If the speed of light depends on the speed of the light source even only in the slightest, then my entire theory of relativity, including the theory of gravity, is wrong.'' (Ref. \refcite{papers1}, doc. 472, p. 555)\footnote{Own translation. Original: ``Sehr neugierig bin ich auch auf die Ergebnisse Ihrer Untersuchungen \"{u}ber die Doppelsterne. Wenn die Lichtgeschwindigkeit auch nur im Geringsten von der Geschwindigkeit der Lichtquelle abh\"{a}ngt, dann ist meine ganze Relativit\"{a}tstheorie inklusive Gravitationstheorie falsch.''}
\end{quote}
In addition to orbital properties, Struble thought that binary systems can provide other observational evidence for special relativity, as we shall discuss now. 

\section{Struble's effect}
\subsection{Motivation and results}
As a student, in the wake of the atomic bombs ending the Second World War, Struble became interested in the underlying physics, the mass defect $\Delta E=\Delta m c^2$ and its origin in special relativity:
\begin{quote}
``Since the constancy [of the speed of light] seemed to me unbelievable, I started thinking about other possible evidence and came upon the double-star phenomena [...].'' (SEC 11)  
\end{quote}
Struble's adviser was initially the Viennese mathematician Karl Menger (1902--1985), a pioneer of fractal and probabilistic metric geometry, who had emigated to the USA joining the University of Notre Dame in 1937, where he started the mathematics PhD program and hosted G\"{o}del. Even after moving to the Illinois Institue of Technology in 1946, he stayed in touch with Struble, whom he had encouraged to work on the binary star problem:
\begin{quote}
``I asked Mr. Struble to study the geometrical aspects of DeSitter's theory and in pursuing this study he discovered another phenomenon which would be observable if the principle [of the constancy of the speed of light] were wrong, namely, a change in the wave length of the light emitted by the companion. This change would be entirely different from the Doppler effect, and quantitatively much larger than DeSitter's phenomena.'' (SEC 4)
\end{quote}
The argument (SEC 3) can be summarized as follows. Consider again a binary star system in emission theory as discussed in Sec. \ref{sec:binary}, with speed of light $c_r$ relative to the observer $E$ given by Eq. (\ref{speed}). Now the observed frequency of light $\nu$ is related to the frequency $\nu_0$ at the emitter $S$ by
\[
\nu=\nu_0 \frac{c_r}{c}\simeq\nu_0 \left(1-\frac{v}{c}\cos\theta\right).
\]
In other words, there is a standard Doppler effect in frequency which is linear in the orbital speed $v$, but it should be noted carefully that the wavelength of light remains unchanged, $\lambda=\lambda_0$, as pointed out by Tolman\cite{tolman1} in 1910. However, Struble observed that the changing of $c_r$ with $\theta$ or emission time $t_0$, that is, the orbital acceleration, gives rise to an actual change in wavelength which can be very large. This is the ``change [...] entirely different from the Doppler effect'' which Menger referred to in the quotation above.

To see this, note that during the time $\Delta t_0=\lambda_0/c$ needed to emit a monochromatic wave of length $\lambda_0$ at $S$, the speed of light relative to $E$ has changed by $\Delta c_r\simeq \Delta t_0 a\omega^2 \sin \omega t_0$, from Eq. (\ref{speed}). So after a travel time of approximately $d/c$ to reach $E$, the wavelength has changed by
\begin{equation}
\Delta \lambda=\lambda-\lambda_0\simeq -\Delta c_r \frac{d}{c}\simeq-\lambda_0 \frac{ad\omega^2}{c^2}\sin \theta
\label{struble}
\end{equation}
at leading order\footnote{This is corrected from Struble's derivation (SEC 3), which contains ambiguities in the angle and sign.}. 
As before in Eq. (\ref{time}), the term containing $d$ may give rise to a large observational effect.

\begin{figure}[H]
\centering
\includegraphics[width=6cm]{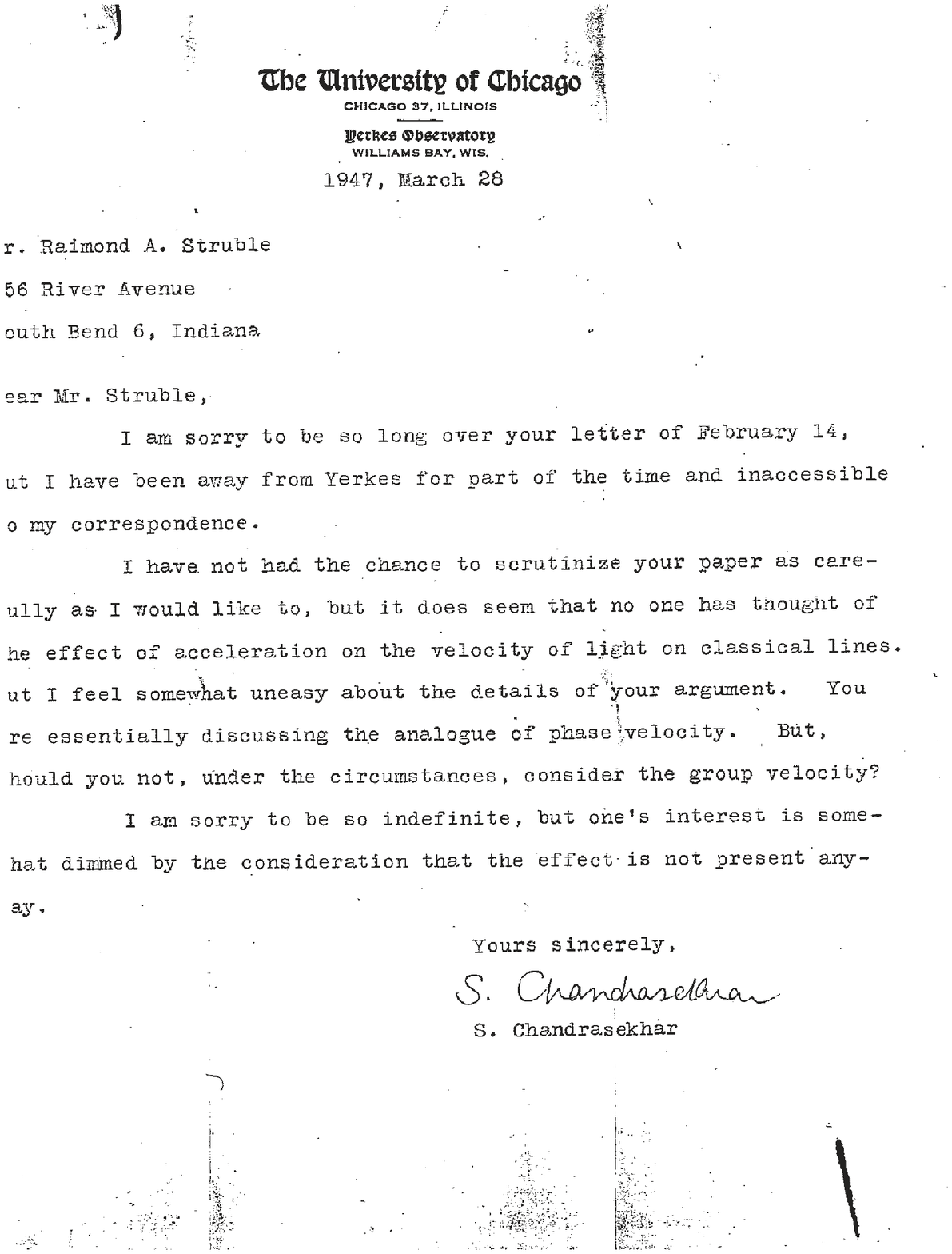}
\caption{Letter by Chandrasekhar to Struble (SEC1).}
\label{fig:chandra}
\end{figure}

In February 1947, Struble expounded his argument in a letter to Chandrasekhar at Yerkes Observatory, who noted in his reply, shown in Fig.~\ref{fig:chandra}, that
\begin{quote}
``[...] it does seem that no one has thought of the effect of acceleration on the velocity of light on classical lines.'' (SEC 1)
\end{quote}
However, he cautioned that group velocity rather than phase velocity should be considered, and that
\begin{quote}
 ``[...] one's interest is somewhat dimmed by the consideration that the effect is not present anyway.'' (SEC 1)
\end{quote}
Special relativity was well established and tested by that time, and the latter remark is also echoed by Tate (SEC 2), the editor of \textit{Physical Review}, where Struble had tried to publish his work as a \textit{Letter to the Editor}. Six days after this rejection, Menger tried intercede on behalf of his former student (SEC 4), and received a reply from the assistant editor Hill (SEC 6). While acknowledging that there are theoretical issues with accelerated light sources even in classical electrodynamics\footnote{This remark presumably refers to the Abraham-Lorentz backreaction force which is proportional to the \textit{third} time derivative of the trajectory, thus causing problems with inital values.}, Hill agreed with Tate that special relativity was already sufficiently corroborated and therefore questioned that
\begin{quote}
``[...] at the present time Ritz' theory requires a final coup de gr\^{a}ce, [...] [having,] like numerous other trial theories of its period, [...] earned a decent oblivion.'' (SEC 6)
\end{quote}
He also pointed out that it remains unclear in Struble's derivation how the spectral energy distribution changes, a remark that reflects Chandrasekhar's concern about the group velocity.

\subsection{Correspondence with Einstein}
Having thus received encouragement as well as rejection, Struble decided to approach Einstein himself as arbiter ``for final judgement'' (SEC 7) regarding the validity of his derivation. In October 1947, Struble sent a letter to Einstein at the Institute for Advanced Study, emphasizing his new effect:
\begin{quote}
``Many spectroscopic binary systems would exhibit these ghost stars due to their geometric characteristics, but as of now I have been unable to find any \underline{visible} binary systems which would. Mostly because of this I investigated geometrical consequences of Ritz' hypothesis which concerned these \underline{visible} double stars, in an attempt to agument [\textit{sic}] de Sitter's repudiation.'' \newline (SEC 7)  
\end{quote}
The accompanying computation (SEC 8) recapitulated the one sketched above (SEC 3), although Struble expressed his result now in terms of frequency rather than wavelength.

\begin{figure}[H]
\centering
\includegraphics[width=12cm,angle=0]{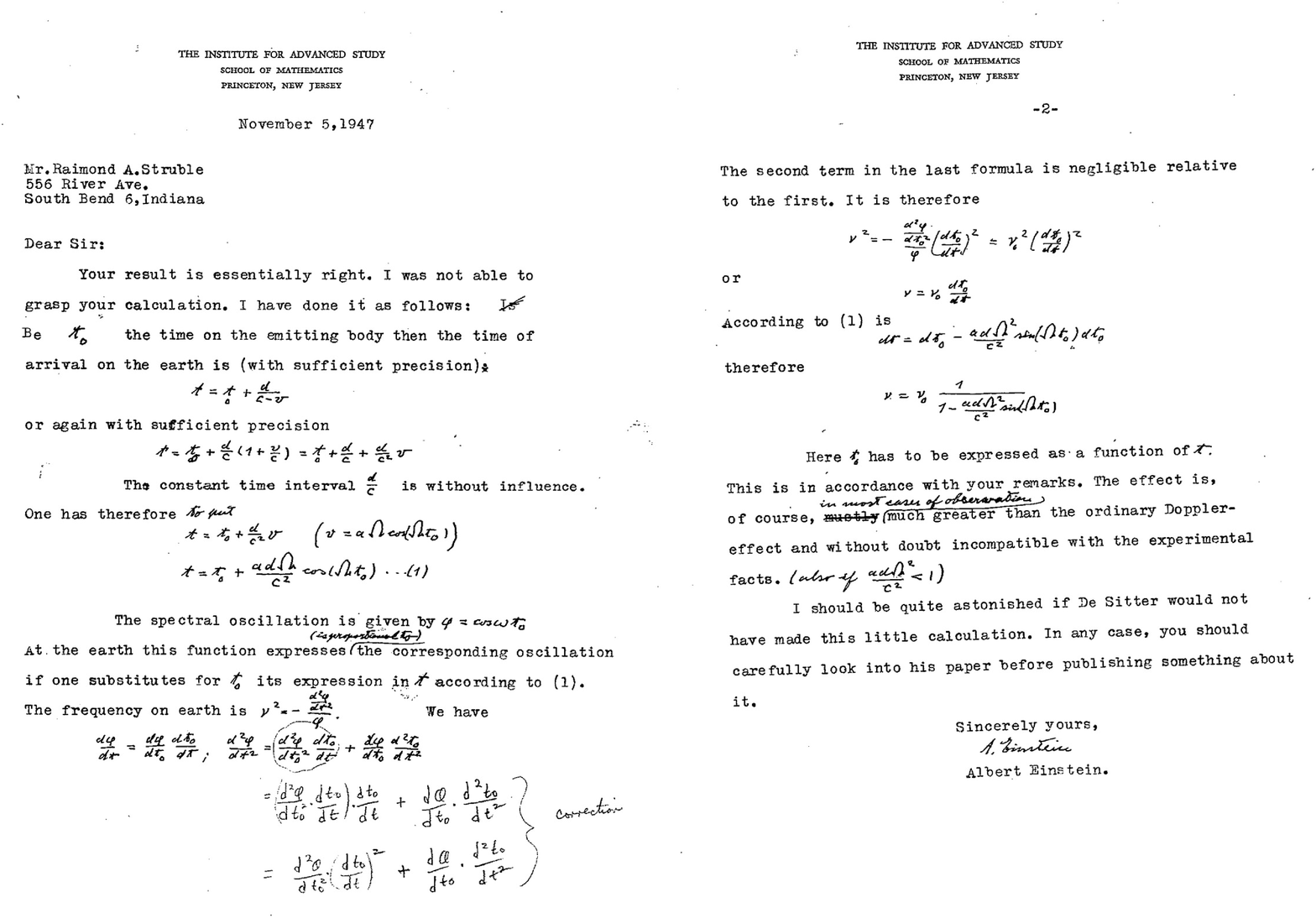}
\caption{Letter by Einstein to Struble (SEC9).}
\label{fig:einstein}
\end{figure}

In his reply of November 1947 (SEC 9), shown in Fig.~\ref{fig:einstein}, Einstein agreed that Struble's argument ``is essentially right'', although he preferred his own derivation. Using again the notation of Sec.~\ref{sec:binary}, Einstein obtained an approximation for the observation time as a function of emission time, 
\[
t\simeq t_0+\frac{ad\omega}{c^2} \cos \omega t_0\,, 
\]
as in Eq. (\ref{time}), ignoring the constant $d/c$ and neglecting the other terms. Infinitesimal intervals are therefore related by
\begin{equation}
\rmd t \simeq \rmd t_0\left(1- \frac{ad\omega^2}{c^2}\sin \omega t_0 \right)\,. 
\label{einstein}
\end{equation}
Assuming that the phase $\varphi$ of the light wave obeys a simple harmonic oscillation, Einstein noted that the frequency is
\[
\nu^2=-\frac{1}{\varphi} \frac{\rmd^2 \varphi}{\rmd t^2}\,,
\]
and the chain rule implies\footnote{The letter has $\frac{\rmd^2 \varphi}{\rmd t^2}=\frac{\rmd^2 \varphi}{\rmd t_0^2}\frac{\rmd t_0}{\rmd t}+\frac{\rmd \varphi}{\rmd t_0}\frac{\rmd ^2 t_0}{\rmd t^2}$, but the mistake in the first term of the right-hand side does not propagate to the following computations. Despite the approximations made, equality signs are used throughout the letter (SEC 9).}
\[
\frac{\rmd^2 \varphi}{\rmd t^2}=\frac{\rmd^2 \varphi}{\rmd t_0^2}\left(\frac{\rmd t_0}{\rmd t}\right)^2+\frac{\rmd \varphi}{\rmd t_0}\frac{\rmd ^2 t_0}{\rmd t^2}\,,
\]
so that, neglecting the last term and using equation (\ref{einstein}), 
\[
\nu\simeq \nu_0\frac{\rmd t_0}{\rmd t}\simeq \frac{\nu_0}{1-\frac{ad\omega^2}{c^2}\sin \omega t_0}\,,
\]
which, in the given approximation, is in agreement with Struble's result. Einstein also supported his interpretation, stating that
\begin{quote}
``The effect is, of course, in most cases of observation much greater than the ordinary Doppler-effect and without doubt incompatible with the experimental facts'' (SEC 9), 
\end{quote}
but, in closing his letter, reminded Struble that careful checking of the literature is in order since
\begin{quote}
``I should be quite astonished if De Sitter would not have made this little calculation. In any case, you should carefully look into his paper before publishing something about it.'' (SEC 9)
\end{quote}

\subsection{Aftermath and assessment}
Einstein's letter made it clear that the question of priority for Struble's result had to be settled before any further attempt at publication should proceed. In January 1948, Menger advised Struble to obtain the original papers regarding the binary star effects of emission theory (SEC 10), in particular the 1913 articles in German by de Sitter\cite{desitter1} and Freundlich\cite{freundlich2}. As discussed in Sec.~\ref{sec:binary}, these are concerned primarily with apparent distortions of the binary orbits -- indeed, Struble's effect is treated neither there nor in the other related papers, by Comstock\cite{comstock1,comstock2}, de Sitter\cite{desitter2,desitter3}, Guthnick\cite{guthnick} and Tolman\cite{tolman1, tolman2}.

A heuristic argument which comes close to Struble's is found in Bergmann's \textit{Introduction to the Theory of Relativity}: using the notation of Sec.~\ref{sec:binary} as before, the light travel time to the observer is approximately $d/c$ so that 
\[
\frac{\Delta t}{\Delta c}\simeq-\frac{d}{c^2}, \quad \mbox {and since} \ \Delta c= v, \quad \Delta t\simeq -\frac{vd}{c^2}\,. 
\]
The amplitude of emission theory effects will then be determined by the ratio $\Delta t/T$, where $T:=2\pi/\omega$ is the orbital period of the binary (Ref. \refcite{bergmann}, p. 20). Hence,
\[
\Delta t \ \omega \simeq -\frac{ad\omega^2}{c^2}\,,
\]
as in Eq. (\ref{struble}) for the amplitude of Struble's effect on wavelength\footnote{In fact, it is likely that Struble knew Bergmann's treatment, since the first edition had been published in 1942 and it had become a popular textbook on relativity theory: there had been four printings by 1948, according to the colophon of the Prentice-Hall, New York NY, 1948, edition. Moreover, Struble used the rather imaginative expression ``ghost stars'' in his letter to Einstein (SEC 7), which is also found in Bergmann (Ref. \refcite{bergmann}, p. 19) but not in the original papers on binary star effects of emission theory mentioned above.}. Still, there is no explicit discussion of Struble's effect and how it differs from the ordinary Doppler effect here.

Hence, it does appear that Struble's effect had not been noticed before or, at any rate, was not at all well known, in accordance with the assessment by Chandrasekhar in SEC 1. It is also interesting to note that the effect does not appear in the 1965 paper by Fox\cite{fox}, which is a detailed critique of all available evidence against emission theories. On the other hand, it must be conceded that Struble's argument is purely kinematical as it stands, and a final assessment of its physical merit should include a deeper investigation of dynamical aspects as well, along the lines suggested by Chandrasekhar (SEC 1) and Hill (SEC 6) mentioned before.

In the end, Struble pursued his interests in mathematics and its applications, and never resumed his attempt to publish this work at the interface of astronomy and theoretical physics:
\begin{quote}
``[...] the main reason I couldn't follow through was that I would need to use Einstein's superior version. Though his letter granted me precidence [\textit{sic}] on the \underline{idea}, I couldn't publish his modifications as mine. I probably should have discussed it with him at the time.'' (SEC 11)
\end{quote}

\section{Final remarks}
In addition to its scientific interest, the Struble--Einstein Correspondence gives detailed insight into individual motivations and practical aspects of the scientific process, as described in this article. Einstein's reply to the student Struble is also a testament to his often generous attitude towards researchers outside his immediate circle of professional colleagues\footnote{Another example of this is his interaction with Mandl in the early history of gravitational lensing, cf. Ref. \refcite{rennsauer}.}

Finally, it may be mentioned that the \textit{Albert Einstein Archives} preserve a copy of Struble's correspondence (SEC 7 in Ref.~\refcite{archive1}, SEC 8 in Ref.~\refcite{archive2}) as well as the autograph draft of Einstein's letter written in German \cite{archive3}, although the English version that Einstein eventually sent to Struble (SEC 9) appears to be lacking.

\section*{Acknowledgments}
MCW would like to thank Masataka Fukugita, Domenico Giulini and Dennis Lehmkuhl for helpful discussions.

\section*{Appendix}
The documents of the Struble-Einstein Correspondence are held by Struble's estate in Raleigh, NC, USA, and consist of the following.
\begin{itemize}
\item[SEC 1.] S. Chandrasekhar to R. A. Struble, typed and signed letter, 1 page, 1947 March 28.
\item[SEC 2.] J. T. Tate to R. A. Struble, typed and signed letter, 1 page, 1947 July 23.
\item[SEC 3.] [R. A. Struble:] An observable consequence of the ballistic hypothesis, typed and annotated document, 1 page (incomplete), no date.
\item[SEC 4.] K. Menger to J. T. Tate, typed letter, 2 pages, 1947 July 29.
\item[SEC 5.] K. Menger [to R. A. Struble], signed autograph letter, 1 page, 1947 August.
\item[SEC 6.] E. L. Hill to K. Menger, typed, signed and annotated letter, 2 pages and copy, 1947 August 11.
\item[SEC 7.] R. A. Struble to A. Einstein, typed letter, 1 page, 1947 October 15.
\item[SEC 8.] R. A. Struble: Consider a double star system [...], typed document, 2 pages, [1947 October 15].
\item[SEC 9.] A. Einstein to R. A. Struble, typed, signed and annotated letter, 2 pages, 1947 November 5.
\item[SEC 10.] [K. Menger] to R. A. Struble, autograph letter, 1 page, 1948 January.
\item[SEC 11.] R. A. Struble, autograph letter, 1 page, 1996 May 5.
\end{itemize}

\newpage

\end{document}